\documentclass[sigconf]{acmart}
\usepackage{graphicx}
\usepackage{subcaption}
\usepackage{float}
\usepackage{geometry}
\AtBeginDocument{%
  }

\copyrightyear{2026}
\acmYear{2026}
\setcopyright{cc}
\setcctype{by}
\acmConference[AHs 2026]{The Augmented Humans International Conference 2026}{March 16--19, 2026}{Okinawa, Japan}
\acmBooktitle{The Augmented Humans International Conference 2026 (AHs 2026), March 16--19, 2026, Okinawa, Japan}
\acmPrice{}
\acmDOI{10.1145/3795011.3797361}
\acmISBN{979-8-4007-2351-3/2026/03}

\begin{document}

\title{Empowering Vocabulary Learning Through Teaching AI: Using LLMs as a Student to Perform Learning by Teaching in Vocabulary Acquisition}

\author{Tokio Uchida}
\authornote{Both authors contributed equally to this research.}
\email{uchida@omu.ac.jp}
\affiliation{%
  \institution{Osaka Metropolitan University}
  \department{Graduate School of Informatics}
  \city{Sakai}
  \state{Osaka}
  \country{Japan}}

\author{Ko Watanabe}
\authornotemark[1]
\email{ko.watanabe@dfki.de}
\orcid{0000-0003-0252-1785}
\affiliation{%
  \institution{DFKI GmbH}
  \city{Kaiserslautern}
  \state{Rheinland-Pfalz}
  \country{Germany}}

\author{Andrew Vargo}
\email{awv@omu.ac.jp}
\affiliation{%
  \institution{Osaka Metropolitan University}
  \department{Graduate School of Informatics}
  \city{Sakai}
  \state{Osaka}
  \country{Japan}
}

\author{Shoya Ishimaru}
\email{ishimaru@omu.ac.jp}
\affiliation{%
  \institution{Osaka Metropolitan University}
  \department{Graduate School of Informatics}
  \city{Sakai}
  \state{Osaka}
  \country{Japan}
}

\author{Ralph L. Rose}
\email{rose@waseda.jp}
\affiliation{%
  \institution{Waseda University}
  \department{Faculty of Science and Engineering}
  \city{Shinjuku}
  \state{Tokyo}
  \country{Japan}}

\author{Ayaka Sugawara}
\email{ayakasug@waseda.jp}
\affiliation{%
  \institution{Waseda University}
  \department{Faculty of Science and Engineering}
  \city{Shinjuku}
  \state{Tokyo}
  \country{Japan}
}

\author{Andreas Dengel}
\email{andreas.dengel@dfki.de}
\affiliation{%
  \institution{DFKI GmbH}
  \city{Kaiserslautern}
  \state{Rheinland-Pfalz}
  \country{Germany}}

\author{Koichi Kise}
\email{kise@omu.ac.jp}
\affiliation{%
  \institution{Osaka Metropolitan University}
  \department{Graduate School of Informatics}
  \city{Sakai}
  \state{Osaka}
  \country{Japan}}

\renewcommand{\shortauthors}{Uchida and Watanabe et al.}

\begin{abstract}
``Learning by Teaching (LbT)'' helps learners deepen their understanding by explaining concepts to others, with questions playing a vital role in identifying knowledge gaps and reinforcing comprehension.
However, existing systems for generating such questions often rely on rigid templates and are expensive to build.
To overcome these limitations, we developed a system using Large Language Models (LLMs) to create dynamic, contextually relevant questions for LbT.
In our English vocabulary learning study, we examined which learner characteristics best leverage the system's benefits.
Our results showed improved memory retention over traditional methods at three and seven days of testing, with ten participants.
Additionally, we identified traits linked to better learning outcomes, highlighting the potential for tailored approaches.
These findings support the development of scalable, cost-effective solutions to enhance LbT methods across various fields.
\end{abstract}



\begin{teaserfigure}
  \centering
  \includegraphics[width=0.9\linewidth]{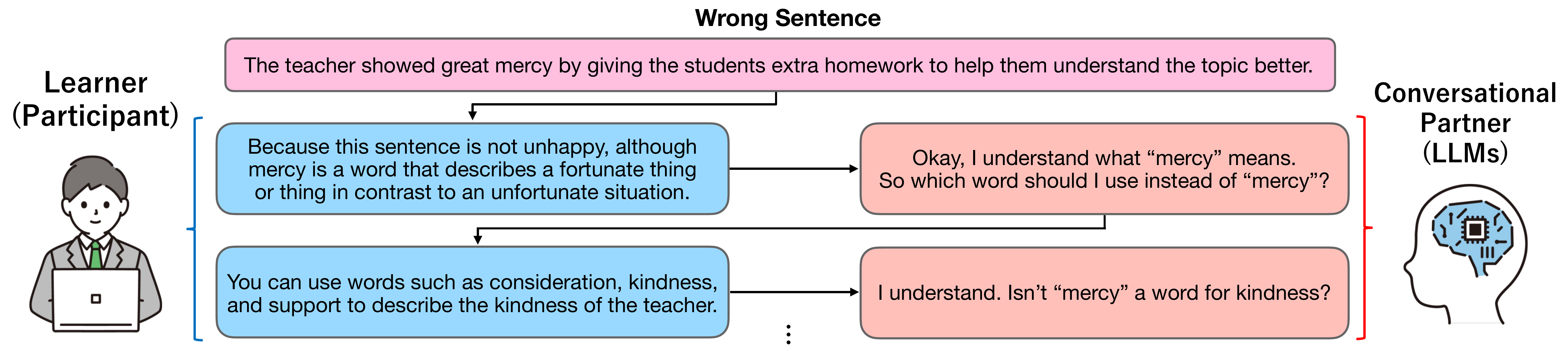}
  \caption{Overview of our proposed system. Our application offer LLMs role as a student to ask questions to English vocabulary learners. Learners enhance learning vocabulary by explaining the definition of the terminology.}
  \label{fig:overview}
\end{teaserfigure}


\maketitle

\section{Introduction}
In the Chinese religious text Liji~\cite{Chik_2021, bookOfRites}, a proverb states, ``Teaching others teaches yourself'', while in Europe, the concept of ``To teach is to learn twice''~\cite{whitman1988peer, weiss1998teach, bowyer2021informal} has been similarly emphasized.
Both ideas, rooted in historical traditions, have been adapted in contemporary Human-Computer Interaction studies under the framework of ``Learning by Teaching (LbT)''~\cite{Nihar24, Amy23, watanabe2026knowledge}.
This educational approach involves learners explaining the learning content to others. 
Research has consistently demonstrated that LbT is highly effective for improving learning outcomes~\cite{Cortese05, Zhu22, Ali24, Maximiliano24}. 
However, the method also presents challenges, such as excessive mental strain, the stress of face-to-face explanations~\cite{Amy23, Wang23}, and the logistical difficulty of finding individuals to interact with as learning partners~\cite{Fiorella13}. 
Developing a system capable of simulating a conversational partner to fulfill this role is essential to addressing these issues and making LbT more accessible and practical.

The advent of Large Language Models (LLMs), particularly GPTs, has revolutionized the substitution of various human roles across domains.
Building on LLMs' exceptional capabilities, this study seeks to leverage these models to simulate a conversational partner in LbT, providing a scalable, interactive solution that overcomes the limitations of traditional human-student setups.
When a learner is asked about a topic they do not fully understand, they can deepen their knowledge by identifying areas that require further study.
Conversely, when asked about a topic they understand well, their motivation and confidence are boosted by correctly answering~\cite{MacIntyre91}.
Thus, the learning effect of LbT is significantly amplified through well-crafted questions.

Previous studies have developed chatbots to generate questions for LbT, but they face significant limitations.
For instance, the questions generated often become formulaic~\cite{Law20, Jin22}, or the systems require extensive source code preparation~\cite{Matsuda22} or predefined, diverse question sets~\cite{Nihar24}, making them costly and time-intensive to build.
These challenges stem from the difficulty of generating meaningful, contextually appropriate questions and accurately measuring their impact on learning effectiveness. 

This research focuses on developing a system that generates questions to promote deeper understanding and meaningful interactions during the teaching process.
\autoref{fig:overview} shows the workflow of LbT in the proposed method.
In this study, we primarily focus on vocabulary acquisition.
English vocabulary acquisition has been one of the significant educational fields of study, which has explored how to enhance learning using smartphones~\cite{higashimura2024estimating}, laptops~\cite{ yamaoka2023experience, yamaoka2025img2vocab}, eye-tracking~\cite{higasa2024keep, ding2023gazereader}, and through interviews~\cite{adjagbodjou2024envisioning, fan2018designing}.
However, the field has yet to evaluate the effectiveness of LbT in vocabulary acquisition. Through iterative question generation and answering, this approach clarifies learners' comprehension and significantly enhances learning outcomes in a scalable, efficient manner.
The research questions (RQs) of this study are as follows:

\begin{enumerate}
\item[RQ1] How can GPT models effectively construct an interactive partner for LbT?
\item[RQ2] What are the interaction patterns and behaviors of learners when engaging with a GPT-simulated conversational partner?
\item[RQ3] To what extent does LbT with LLMs enhance vocabulary acquisition?
\end{enumerate}

\section{Related Work}
In this section, we introduce studies investigating the learning effects of LbT and studies that have used LLMs to construct student models. 
We also describe the relationship between these studies and our work.

LbT is a method of deepening knowledge by teaching what learners have learned to others.
This learning method is not only effective when applied to members of the same school or organization~\cite{Cortese05, Maximiliano24}, but also when applied to Chatbots~\cite{Jin22, Jin24, Matsuda22, Nihar24} or simply talking to an animated image of a person~\cite{Zhu22}.
LbT is especially enhanced by actually teaching others~\cite{Fiorella13}.
However, it has also been shown that psychological barriers to learning exist.
\citet{Wang23} have university students who attended a lecture do three different types of LbT, and the results showed that making a video was the most effective because of excessive mental strain and stress of explaining to a face-to-face person. 
\citet{Amy23} also conducted an experiment in which college students practiced LbT, and the results showed that students also experienced psychological barriers. While LbT is highly effective for learning, there are disadvantages to using it on an actual human. This research aims to build a system that acts as a student, rather than people, to achieve more effective LbT.

In recent years, LLMs have been developed and used in education. 
For instance, an interactive chat tool exists for educational interns to train~\cite{Julia23}. 
This tool is based on GPT-3 and uses prompt engineering to construct a pseudo-student model with a unique personality that can interact as naturally as a real student.
In addition, AI agents have been developed to promote student participation in VR lessons~\cite {Liu24}.
This agent is based on GPT-4 and has been built as a 3D avatar that intervenes in lessons based on the situation of surrounding students and the lecture through prompt engineering.
Using LLMs in this way enables the construction of systems that can generate natural-sounding sentences similar to those produced by humans. 

\section{Proposed System}
In this study, the learning content of LbT is the acquisition of English vocabulary and idioms. 
We focus on acquiring English vocabulary and idioms, which are the foundation of English language learning.
As teaching materials for LbT, we use English sentences and words or idioms that the learner does not understand well.
As part of the LbT task, the learner corrects English sentences by replacing incorrect words or idioms with correct ones and explains the reasons for these corrections.
Through this task, the learner can deepen their understanding of English words and idioms.
We generate these teaching materials using GPT-4o based on the prompts (\autoref{appendix:prompts}).
When given an English word or idiom, LLMs use the prompt to output a sentence with errors, the reasons, and a corrected example.

Our system generates questions using GPT-4o.
This model is characterized by its ability to generate sentences both quickly and accurately compared to GPT-4 and GPT-4-turbo.
Existing studies have demonstrated that dialogue systems can be constructed using prompt engineering based on LLMs~\cite{Julia23, Liu24}, and that strong learning effects can be achieved by conducting LbT with a conversational partner who has less knowledge than the learner~\cite{Ali24}.
Therefore, we designed the system to simulate a beginner in English by instructing GPT-4o through prompts.

In the learning flow of the proposed system, the learner first reads an English sentence that contains a word or idiom they are unfamiliar with and misuse. 
The learner then starts LbT by explaining the reasons for the correction.
When the learner inputs their explanation into the system, LLMs generate appropriate questions based on the wrong sentence and the provided explanation. 
The learner then creates an answer to the generated question and re-enters it into the system.
By repeating the process in which LLMs generate questions and learners provide answers, we aim to implement the concept of LbT.

\section{Experiment}
We experimented with ten university students.
As an overview of the experiment, we used the proposed system with LLMs as the experimental system, and the other without LLMs as the baseline, and changes in understanding before and after learning were measured using Pre-Post-tests.
During the experiment, a questionnaire was also administered to the participants, asking about their impressions of the study and the difficulty level of the test.
The participants were rewarded with 5,000 yen for a five-hour experiment.
As a Pretest to identify English words and idioms that the learners did not understand well, we prepared multiple-choice questions that ask about vocabulary or idioms from past Eiken tests~\footnote{\url{https://www.eiken.or.jp/eiken/}} of the appropriate grade level for the learners' English ability. 
As for the Posttests, we generated multiple-choice questions on the vocabulary and idioms tested in the Pretest using GPT-4o.



The participants first answered a Pretest consisting of 30 multiple-choice questions on vocabulary.
Next, they studied the words and idioms they got wrong in the Pretest, either learning with the proposed or baseline system. 
Immediately after, they answered another multiple-choice question on ten of the 30 vocabulary items from the Pretest as Posttest-1 (\autoref{appendix:questionaires}).
They then did another Pretest on 30 vocabulary items, studied the words they got wrong using the other system than the one they used before, and did Posttest-1. 
Then, for the remaining 20 words studied using each system, they took Posttest-2 and Posttest-3, each with ten words, three days and seven days later, respectively. 
To avoid order effects in this experiment, the participants were divided into Group A and Group B. Half started by learning using the proposed system, while the other half started using the baseline system. 
The questionnaire was administered before the experiment began after each Posttest-1 and after the two Posttests.

This experiment was conducted using a web application developed by Vite+React and FastAPI on a personal computer owned by each participant.
Participants answered the multiple-choice questions on the screen (\autoref{appendix:pretest}).
For every ten questions answered, the answer results and explanations (only for the Pretest) were displayed.
The maximum number of corrections for incorrect English sentences was set to five. If the learner could not correct the sentence by the fifth correction, the correct correction was displayed. 
The time for conducting LbT was set to three minutes, and the first question was fixed with ``Please explain the reasons for the corrections.'' 
Although the use of dictionaries and translation devices was prohibited during the experiment, the meanings of words were displayed on the screen during the study. 
As for the language used within the application, participants were asked to select their native language (Japanese or English) when logging in to the application, and the explanations of the questions were displayed in the selected language. 
LbT was also carried out in each participant's native language. 

\begin{figure}[t]
  \centering
  \includegraphics[width=\columnwidth]{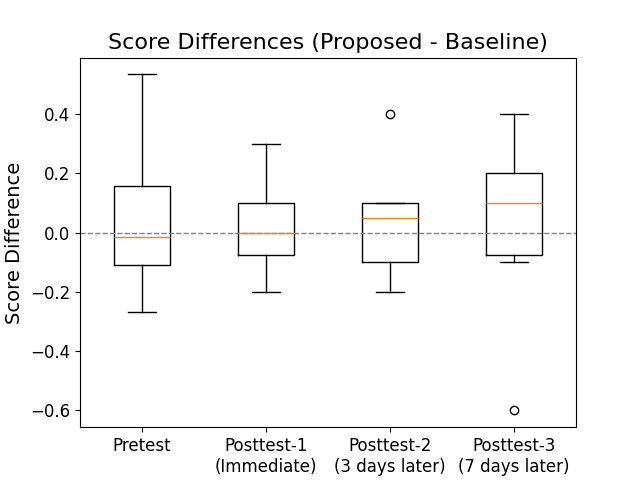}
  \caption{Difference between the percentage of correct answers in each test using the proposed system and the baseline.}
  \label{fig:scoreBox}
\end{figure}

\begin{figure}[t]
  \centering
  \includegraphics[width=\columnwidth]{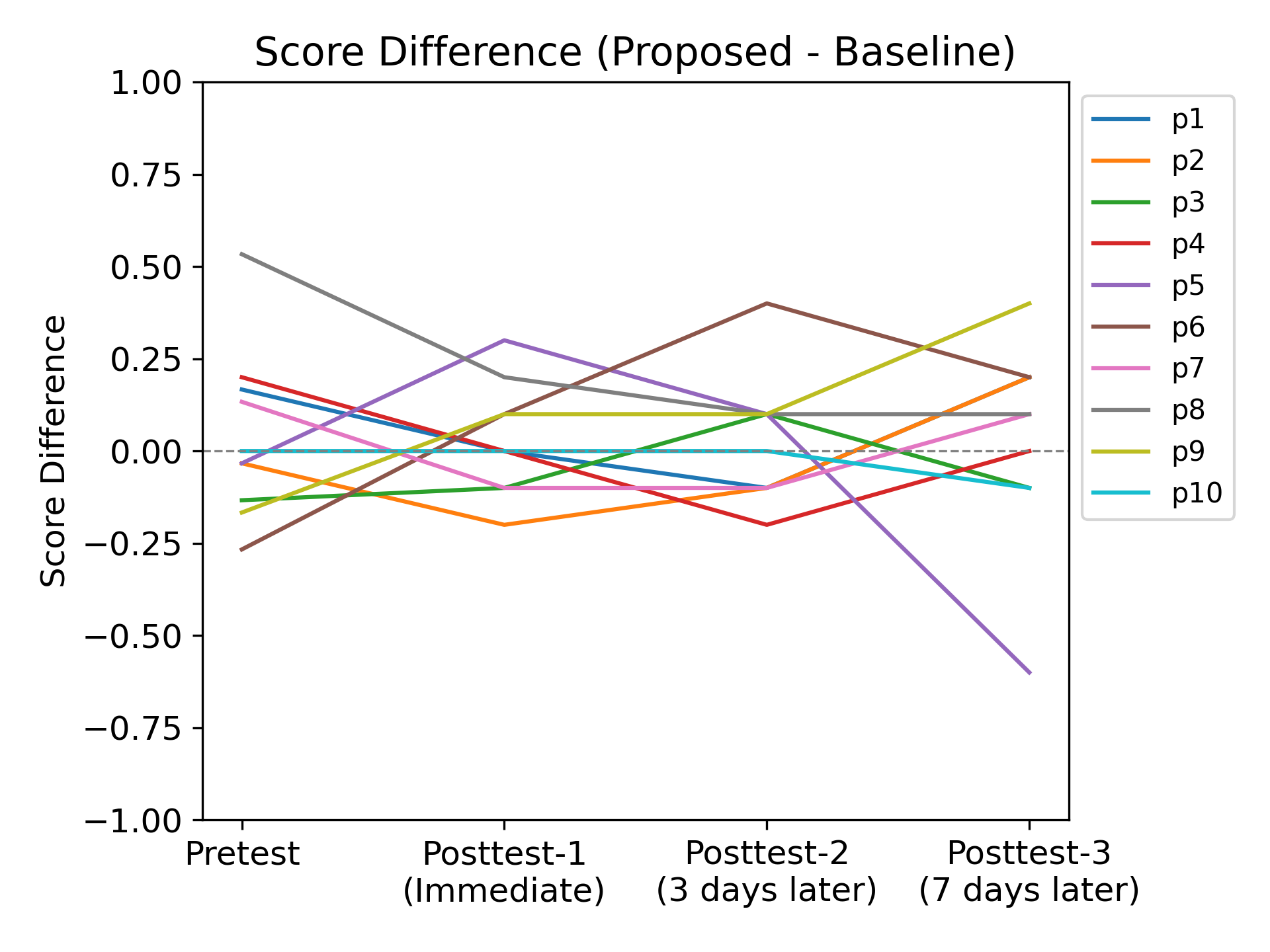}
  \caption{Difference in the percentage of correct answers during learning between the proposed and baseline systems.}
  \label{fig:scoreBar}
\end{figure}

\section{Results and Discussion}
\autoref{fig:scoreBox} shows the score (percentage of correct answers) difference between the proposed system and that when using the baseline system.
The distributions of Pretest and Posttest-1 are more centered, while the distributions shift positively over Posttests-2 and 3, suggesting high memory retention.
Despite the overall positive outcomes, individual differences in learning effectiveness were observed. 

\autoref{fig:scoreBar} presents line graphs illustrating the percentage of correct answers for each test, calculated as the difference between the percentage of correct answers with the proposed system and those with the baseline system across participants. 
While the majority benefited from the proposed system, some cases highlight challenges in adapting to it. 
For instance, participant p5's percentage of correct answers is 60\% lower when learning with the proposed system than when learning with the baseline system. 
The leading cause of this is his overall learning experience. Feedback from p5 indicated that engaging with the system was ``stressful,'' likely due to cognitive overload from trying to learn too many words at once.

Participants such as p2 and p6, who entered more words per interaction, achieved higher learning outcomes (\autoref{appendix:inquiryWordsAverage}). 
However, participants like p5, who both entered fewer words, showed significantly reduced learning gains. 
As for the participant who showed learning effects despite the small number of letters he typed (p9), he stated in the questionnaire, ``I feel like my memory has been retained after typing many times.'' 
It may be because the participant learned without feeling overloaded.
These findings emphasize the importance of active engagement and tailored learning approaches when interacting with LLMs. 
Participants who submitted detailed, well-thought-out inputs benefited more from the system.
Additionally, the cognitive effort required during learning must align with the learner's capacity and prior knowledge to avoid overloading them.

In summary, the proposed system can significantly enhance vocabulary learning and retention, provided that learners actively engage in meaningful interactions with the LLMs.
Moreover, balancing the number of words studied and tailoring the learning process to individual needs is crucial for maximizing learning outcomes.
Future work could explore adaptive systems that dynamically adjust to a learner's input style and cognitive capacity to further optimize the learning experience.

Our study demonstrates that learning with the proposed system results in high learning effectiveness.
When asked about their impressions of the dialogue content with LLMs, several participants mentioned that they were repeatedly asked the same questions and could predict the next question.
We then found cases where the same questions were asked repeatedly (\autoref{appendix:badExample}). 
This issue may stem from the model's deterministic, repetitive output, as the temperature hyperparameter of GPT-4o was set to zero, and the prompts did not strongly emphasize adapting to the learner's responses. 


\section{Conclusion}
In this study, we aim to implement LbT by constructing a conversational partner using GPT-4o. The results demonstrate that our proposed question-generation system enhances memory retention in English learning. Our results showed improved memory retention at 3 and 7 days after learning. Our findings show that participants who engaged in communication with GPT as a conversational partner achieved better results in English vocabulary acquisition. In future work, we plan to improve the system further, increase the number of participants, and conduct additional experiments to make LbT more effective.



\bibliographystyle{ACM-Reference-Format}
\bibliography{main}

@String{Computing = "Computing" }

@String{Computer = "{IEEE} Computer" }

@String{Springer = "Springer-Verlag" }

@article{Maximiliano24,
  author  = {Paredes-Velasco, Maximiliano and Lozano-Osorio, Isaac and Pérez-Marín, Diana and Santacruz-Valencia, Liliana Patricia},
  title   = {A Case Study on Learning Visual Programming With TutoApp for Composition of Tutorials: An Approach for Learning by Teaching},
  journal = {IEEE Transactions on Learning Technologies},
  volume  = {17},
  number  = {},
  month   = {},
  year    = {2024},
  pages   = {498--513},
  doi     = {10.1109/TLT.2022.3226122},
  url     = {},
  note    = {}
}

@article{Amy23,
  author  = {Debban\'{e}, Amy and Lee, Ken Jen and Tse, Jarvis and Law, Edith},
  title   = {Learning by Teaching: Key Challenges and Design Implications},
  journal = {Proc. ACM Hum.-Comput. Interact.},
  volume  = {7},
  number  = {CSCW1},
  month   = apr,
  year    = {2023},
  pages   = {1--34},
  doi     = {10.1145/3579501},
  url     = {https://doi.org/10.1145/3579501},
  note    = {}
}

@article{Wang23,
  author  = {Wang, Fuxing and Cheng, Meixia and Mayer, Richard},
  title   = {Improving learning-by-teaching without audience interaction as a generative learning activity by minimizing the social presence of the audience.},
  journal = {Journal of Educational Psychology},
  volume  = {115},
  number  = {6},
  month   = {},
  year    = {2023},
  pages   = {783--797},
  doi     = {10.1037/edu0000801},
  url     = {https://doi.org/10.1037/edu0000801},
  note    = {}
}

@article{Matsuda22,
  author  = {Matsuda, Noboru},
  title   = {Teachable Agent as an Interactive Tool for Cognitive Task Analysis: A Case Study for Authoring an Expert Model},
  journal = {International Journal of Artificial Intelligence in Education},
  volume  = {32},
  number  = {},
  month   = {},
  year    = {2022},
  pages   = {48--75},
  doi     = {10.1007/s40593-021-00265-z},
  url     = {https://doi.org/10.1007/s40593-021-00265-z},
  note    = {}
}

@article{Fiorella13,
  author  = {Logan Fiorella and Richard E. Mayer},
  title   = {The relative benefits of learning by teaching and teaching expectancy},
  journal = {Contemporary Educational Psychology},
  volume  = {38},
  number  = {4},
  month   = {},
  year    = {2013},
  pages   = {281--288},
  doi     = {https://doi.org/10.1016/j.cedpsych.2013.06.001},
  url     = {https://www.sciencedirect.com/science/article/pii/S0361476X13000209},
  note    = {}
}

@article{Cortese05,
  author  = {Claudio G. Cortese},
  title   = {Learning through Teaching},
  journal = {Management Learning},
  volume  = {36},
  number  = {1},
  month   = {},
  year    = {2005},
  pages   = {87--115},
  doi     = {10.1177/1350507605049905},
  url     = {https://doi.org/10.1177/1350507605049905},
  note    = {}
}

@article{MacIntyre91,
  author  = {Gardner, R. C. and MacIntyre, P. D.},
  title   = {An Instrumental Motivation In Language Study: Who Says It Isn’t Effective?},
  journal = {Studies in Second Language Acquisition},
  volume  = {13},
  number  = {1},
  month   = {},
  year    = {1991},
  pages   = {57--72},
  doi     = {10.1017/S0272263100009724},
  url     = {},
  note    = {}
}

@inproceedings{Ali24,
  author       = {Malik, Ali and Woodrow, Juliette and Piech, Chris},
  title        = {Learners Teaching Novices: An Uplifting Alternative Assessment},
  booktitle    = {Proceedings of the 55th ACM Technical Symposium on Computer Science Education V. 1},
  series       = {SIGCSE 2024},
  editor       = {},
  volume       = {},
  year         = {2024},
  publisher    = {Association for Computing Machinery},
  address      = {New York, NY, USA},
  pages        = {785–-791},
  doi          = {10.1145/3626252.3630887},
  url          = {https://doi.org/10.1145/3626252.3630887},
  number       = {},
  month        = {},
  organization = {},
  note         = {}
}

@inproceedings{Jin24,
  author       = {Jin, Hyoungwook and Lee, Seonghee and Shin, Hyungyu and Kim, Juho},
  title        = {Teach AI How to Code: Using Large Language Models as Teachable Agents for Programming Education},
  booktitle    = {Proceedings of the 2024 CHI Conference on Human Factors in Computing Systems},
  series       = {CHI '24},
  editor       = {},
  volume       = {},
  year         = {2024},
  publisher    = {Association for Computing Machinery},
  address      = {New York, NY, USA},
  pages        = {1–-28},
  doi          = {10.1145/3613904.3642349},
  url          = {https://doi.org/10.1145/3613904.3642349},
  number       = {},
  month        = {},
  organization = {},
  note         = {}
}

@inproceedings{Liu24,
  author       = {Liu, Ziyi and Zhu, Zhengzhe and Zhu, Lijun and Jiang, Enze and Hu, Xiyun and Peppler, Kylie A and Ramani, Karthik},
  title        = {ClassMeta: Designing Interactive Virtual Classmate to Promote VR Classroom Participation},
  booktitle    = {Proceedings of the 2024 CHI Conference on Human Factors in Computing Systems},
  series       = {CHI '24},
  editor       = {},
  volume       = {},
  year         = {2024},
  publisher    = {Association for Computing Machinery},
  address      = {New York, NY, USA},
  pages        = {1–-17},
  doi          = {10.1145/3613904.3642947},
  url          = {https://doi.org/10.1145/3613904.3642947},
  number       = {},
  month        = {},
  organization = {},
  note         = {}
}

@inproceedings{Nihar24,
  author       = {Sabnis, Nihar and Nagashima, Tomohiro},
  title        = {Empowering Learners: Chatbot-Mediated 'Learning-by-Teaching'},
  booktitle    = {Extended Abstracts of the CHI Conference on Human Factors in Computing Systems},
  series       = {CHI EA '24},
  editor       = {},
  volume       = {},
  year         = {2024},
  publisher    = {Association for Computing Machinery},
  address      = {New York, NY, USA},
  pages        = {1–-9},
  doi          = {10.1145/3613905.3650754},
  url          = {https://doi.org/10.1145/3613905.3650754},
  number       = {},
  month        = {},
  organization = {},
  note         = {}
}

@inproceedings{Julia23,
  author       = {Markel, Julia M. and Opferman, Steven G. and Landay, James A. and Piech, Chris},
  title        = {GPTeach: Interactive TA Training with GPT-based Students},
  booktitle    = {Proceedings of the Tenth ACM Conference on Learning @ Scale},
  series       = {L@S '23},
  editor       = {},
  volume       = {},
  year         = {2023},
  publisher    = {Association for Computing Machinery},
  address      = {New York, NY, USA},
  pages        = {226–-236},
  doi          = {10.1145/3573051.3593393},
  url          = {https://doi.org/10.1145/3573051.3593393},
  number       = {},
  month        = {},
  organization = {},
  note         = {}
}

@inproceedings{Zhu22,
  author       = {Zhu, Fangfang and Yang, Jiumin and Pi, Zhongling},
  title        = {Benefits of Peer Learning and Learning by Teaching for Students Learning through Instructional Videos},
  booktitle    = {2022 IEEE 2nd International Conference on Educational Technology (ICET)},
  series       = {},
  editor       = {},
  volume       = {},
  year         = {2022},
  publisher    = {},
  address      = {},
  pages        = {96--100},
  doi          = {10.1109/ICET55642.2022.9944478},
  url          = {},
  number       = {},
  month        = {},
  organization = {},
  note         = {}
}

@inproceedings{Jin22,
  author       = {Jin, Nayoung  and Lee, Hana},
  title        = {StuBot: Learning by Teaching a Conversational Agent Through Machine Reading Comprehension},
  booktitle    = {Findings of the Association for Computational Linguistics: EMNLP 2022},
  series       = {},
  editor       = {Goldberg, Yoav  and Kozareva, Zornitsa  and Zhang, Yue},
  volume       = {},
  year         = {2022},
  publisher    = {Association for Computational Linguistics},
  address      = {Abu Dhabi, United Arab Emirates},
  pages        = {3008--3020},
  doi          = {10.18653/v1/2022.findings-emnlp.219},
  url          = {https://aclanthology.org/2022.findings-emnlp.219},
  number       = {},
  month        = dec,
  organization = {},
  note         = {}
}

@inproceedings{Law20,
  author       = {Law, Edith and Baghaei Ravari, Parastoo and Chhibber, Nalin and Kulic, Dana and Lin, Stephanie and Pantasdo, Kevin D. and Ceha, Jessy and Suh, Sangho and Dillen, Nicole},
  title        = {Curiosity Notebook: A Platform for Learning by Teaching Conversational Agents},
  booktitle    = {Extended Abstracts of the 2020 CHI Conference on Human Factors in Computing Systems},
  series       = {CHI EA '20},
  editor       = {},
  volume       = {},
  year         = {2020},
  publisher    = {Association for Computing Machinery},
  address      = {New York, NY, USA},
  pages        = {1--9},
  doi          = {10.1145/3334480.3382783},
  url          = {https://doi.org/10.1145/3334480.3382783},
  number       = {},
  month        = {},
  organization = {},
  note         = {}
}

@misc{Chik_2021,
  series     = {UBC Community and Partner Publications},
  title      = {Liji (Book of Rites)},
  url        = {https://open.library.ubc.ca/collections/ubccommunityandpartnerspublicati/52387/items/1.0404466},
  doi        = {http://dx.doi.org/10.14288/1.0404466},
  publisher  = {Database of Religious History (DRH)},
  author     = {Chik, HinMingFrankie},
  year       = {2021},
  month      = {Sep},
  collection = {UBC Community and Partner Publications}
}

@article{weiss1998teach,
  title     = {To teach is to learn twice: resident teachers learn more},
  author    = {Weiss, Victoria and Needlman, Robert},
  journal   = {Archives of pediatrics \& adolescent medicine},
  volume    = {152},
  number    = {2},
  pages     = {190--192},
  year      = {1998},
  publisher = {American Medical Association}
}

@article{bowyer2021informal,
  title     = {Informal near-peer teaching in medical education: a scoping review},
  author    = {Bowyer, Eleanor R and Shaw, Sebastian CK},
  journal   = {Education for Health},
  volume    = {34},
  number    = {1},
  pages     = {29--33},
  year      = {2021},
  publisher = {Medknow}
}

@book{bookOfRites,
  title      = {The Book of Rites (Liji): Bilingual Edition, English and Chinese},
  author     = {{Confucius (Attributed)}},
  translator = {James Legge},
  year       = {2014},
  publisher  = {James Legge},
  isbn       = {978-1500581023},
  url        = {https://www.amazon.de/Book-Rites-Liji-Bilingual-English-ebook/dp/B00KVGYS9M},
  note       = {Bilingual Edition, English and Chinese}
}

@book{whitman1988peer,
  title     = {Peer Teaching: To Teach Is To Learn Twice. ASHE-ERIC Higher Education Report No. 4, 1988.},
  author    = {Whitman, Neal A and Fife, Jonathan D},
  year      = {1988},
  publisher = {ERIC}
}

@article{higashimura2024estimating,
  author   = {Higashimura, Riku and Watanabe, Ko and Vargo, Andrew and Iwata, Motoi and Dengel, Andreas and Kise, Koichi},
  journal  = {IEEE Access},
  title    = {Estimating Unknown English Words From User Smartphone Reading Behaviors},
  year     = {2024},
  volume   = {12},
  number   = {},
  pages    = {140223-140234},
  doi      = {10.1109/ACCESS.2024.3457510}
}

@inproceedings{yamaoka2023experience,
  author    = {Yamaoka, Kanta and Watanabe, Ko and Kise, Koichi and Dengel, Andreas and Ishimaru, Shoya},
  title     = {Experience is the Best Teacher: Personalized Vocabulary Building Within the Context of Instagram Posts and Sentences from GPT-3},
  year      = {2023},
  isbn      = {9781450394239},
  publisher = {Association for Computing Machinery},
  address   = {New York, NY, USA},
  url       = {https://doi.org/10.1145/3544793.3560382},
  doi       = {10.1145/3544793.3560382},
  booktitle = {Adjunct Proceedings of the 2022 ACM International Joint Conference on Pervasive and Ubiquitous Computing and the 2022 ACM International Symposium on Wearable Computers},
  pages     = {313–316},
  numpages  = {4},
  location  = {Cambridge, United Kingdom},
  series    = {UbiComp/ISWC '22 Adjunct}
}

@inproceedings{higasa2024keep,
  author    = {Higasa, Taichi and Tanaka, Keitaro and Feng, Qi and Morishima, Shigeo},
  title     = {Keep Eyes on the Sentence: An Interactive Sentence Simplification System for English Learners Based on Eye Tracking and Large Language Models},
  year      = {2024},
  isbn      = {9798400703317},
  publisher = {Association for Computing Machinery},
  address   = {New York, NY, USA},
  url       = {https://doi.org/10.1145/3613905.3650792},
  doi       = {10.1145/3613905.3650792},
  booktitle = {Extended Abstracts of the CHI Conference on Human Factors in Computing Systems},
  articleno = {211},
  numpages  = {7},
  location  = {Honolulu, HI, USA},
  series    = {CHI EA '24}
}

@inproceedings{adjagbodjou2024envisioning,
  author    = {Adjagbodjou, Adinawa and Kaufman, Geoff},
  title     = {Envisioning Support-Centered Technologies for Language Practice and Use: Needs and Design Opportunities for Immigrant English Language Learners (ELLs)},
  year      = {2024},
  isbn      = {9798400703300},
  publisher = {Association for Computing Machinery},
  address   = {New York, NY, USA},
  url       = {https://doi.org/10.1145/3613904.3642236},
  doi       = {10.1145/3613904.3642236},
  booktitle = {Proceedings of the 2024 CHI Conference on Human Factors in Computing Systems},
  articleno = {568},
  numpages  = {15},
  location  = {Honolulu, HI, USA},
  series    = {CHI '24}
}

@inproceedings{ding2023gazereader,
  author    = {Ding, Jiexin and Zhao, Bowen and Huang, Yuqi and Wang, Yuntao and Shi, Yuanchun},
  title     = {GazeReader: Detecting Unknown Word Using Webcam for English as a Second Language (ESL) Learners},
  year      = {2023},
  isbn      = {9781450394222},
  publisher = {Association for Computing Machinery},
  address   = {New York, NY, USA},
  url       = {https://doi.org/10.1145/3544549.3585790},
  doi       = {10.1145/3544549.3585790},
  booktitle = {Extended Abstracts of the 2023 CHI Conference on Human Factors in Computing Systems},
  articleno = {149},
  numpages  = {7},
  location  = {Hamburg, Germany},
  series    = {CHI EA '23}
}

@inproceedings{fan2018designing,
  author    = {Fan, Min and Jin, Sheng and Antle, Alissa N.},
  title     = {Designing Colours and Materials in Tangible Reading Products for Foreign Language Learners of English},
  year      = {2018},
  isbn      = {9781450356213},
  publisher = {Association for Computing Machinery},
  address   = {New York, NY, USA},
  url       = {https://doi.org/10.1145/3170427.3188577},
  doi       = {10.1145/3170427.3188577},
  booktitle = {Extended Abstracts of the 2018 CHI Conference on Human Factors in Computing Systems},
  pages     = {1–6},
  numpages  = {6},
  location  = {Montreal QC, Canada},
  series    = {CHI EA '18}
}

@article{yamaoka2025img2vocab,
  author   = {Yamaoka, Kanta and Watanabe, Ko and Kise, Koichi and Dengel, Andreas and Ishimaru, Shoya},
  journal  = {IEEE Access},
  title    = {Img2Vocab: Explore Words Tied to Your Life with LLMs and Social Media Images},
  year     = {2025},
  volume   = {},
  number   = {},
  pages    = {1-1},
  doi      = {10.1109/ACCESS.2025.3533076}
}

@inproceedings{watanabe2026knowledge,
  title        = {Knowledge Transfer with AI},
  author       = {Watanabe, Ko and Gro{\ss}mann, Nicolas and Maerz, Christoph and Ishimaru, Shoya and Dengel, Andreas},
  booktitle    = {The Future of Education with AI: Communications of NII Shonan Meetings},
  pages        = {51--86},
  year         = {2026},
  organization = {Springer}
}

\newpage
\appendix

\section{Prompts used in the experiment}
\label{appendix:prompts}

\begin{figure}[h]
  \centering
  \small
  \begin{tabular}{|p{\dimexpr\linewidth-2\tabcolsep-2\arrayrulewidth\relax}|}
    \hline
    You are an expert in English language learning. \\
    Please make a wrong English sentence including the given keyword. \\
    Please follow these notes when making the sentence. \\ 
    \\
    \# Notes \\
    - You will be given a keyword. \\
    - The sentence must include the given keyword. \\
    - You must use incorrectly but plausibly the given keyword in the sentence. \\
    - The length of the sentence must be around 30 words. \\
    - You must also provide a detailed reason why the sentence is incorrect. \\
    - You must also provide a list of words with the incorrect keyword corrected to correct ones. \\
    - When there are multiple correct words, you must provide all of them. \\
    - The output must follow the JSON format below. \\ 
    \\
    \# JSON format \\
    \{ \\
    \quad "title": Please follow the format below: Misuse of the "(keyword)", \\
    \quad "content": Incorrect English sentence including the given keyword, \\
    \quad "evidence": Reason why the sentence is incorrect in detail, \\
    \quad "conversion": [ \\
    \qquad \{ \\
    \qquad \quad "incorrect": Keyword which is used incorrectly, \\
    \qquad \quad "correct": One of the word correcting the keyword used incorrectly \\
    \qquad \} \\
    \qquad ... \\
    \quad ] \\
    \} \\
    \hline
  \end{tabular}
  \caption{Prompt feed to GPT-4o to generate material for sentence correction. GPT-4o generates a sentence with a misused keyword.}
  \label{fig:material_prompt}
\end{figure}

\begin{figure}[h]
  \centering
  \small
  \begin{tabular}{|p{\dimexpr\linewidth-2\tabcolsep-2\arrayrulewidth\relax}|}
    \hline
    You are a student taking an English class. \\
    Please follow the settings and instructions to create a question. \\
    \\
    \# Settings \\
    - Your characteristics as a student are as follows: \\
    \quad - You are a beginner English learner. \\
    \quad - You have a habit of putting on an agreement, such as ``I see'' or ``I understand'' before asking a question. \\
    - Your English proficiency is as follows:. \\
    \quad - Vocabulary: Knows words and phrases for basic everyday conversation, but vocabulary is limited. \\
    \quad - Grammar: The student has begun to understand basic grammatical rules, but there are still many inaccuracies. \\
    \quad - Reading Comprehension: Can understand basic sentences and short paragraphs, but at a slower rate. \\
    \quad - Writing: Can write simple sentences with many errors. \\ 
    \\
    \# Instructions \\
    - The learning material for the class is the following sentence: {material['content']}. \\
    - In the sentence, the word "{keyword}" is used incorrectly. \\
    - Yow will be given the following information by your teacher: \\
    \quad - You will be given a question and an explanation of the mistake in the sentence from your teacher. \\
    \quad - You need to make another question based on the question and explanation given by the teacher. \\
    \quad - The question you make must be related to the student's characteristics regarding English proficiency. \\
    - You must not create the same inquiry as the previous inquiries. \\
    \quad - The previous inquiries are as follows: \{", ".join(recent\_inquiries)\}. \\
    \\
    \# Notes \\
    - Output should be the question only. \\
    - Use a variety of types of agreements in addition to those listed in the examples. \\
    - Make the output as natural as if an actual student were speaking. \\
    - The question must be written in {language}. But the keyword should be written in English. \\
    \hline
  \end{tabular}
  \caption{Prompt fed to GPT-4o to simulate a beginner-level conversational partner.}
  \label{fig:student_prompt}
\end{figure}



\clearpage
\section{Questionaires used in the experiment}
\label{appendix:questionaires}

\begin{table}[h]
  \centering
  \small
  \caption{List of questions before the experiment}
  \label{tab:survey1}
  \begin{tabular}{c|p{\dimexpr\columnwidth-4em-2\tabcolsep-2\arrayrulewidth\relax}}
  \hline
         & Survey Questions \\ \hline
    Q1 & What is your name? \\
    Q2 & What is your age? \\
    Q3 & What is your gender? \\
    Q4 & How much English do you use in your daily life? \\
    Q5 & Have you ever practiced Learning by Teaching (learning to teach others what you learned)? \\
  \hline
  \end{tabular}
\end{table}

\begin{table}[h]
  \centering
  \small
  \caption{List of questions after learning with the proposed system}
  \label{tab:survey2}
  \begin{tabular}{c|p{\dimexpr\columnwidth-4em-2\tabcolsep-2\arrayrulewidth\relax}}
  \hline
         & Survey Questions \\ \hline
    Q1 & How difficult was the Pretest? \\
    Q2 & How difficult was the Posttest? \\
    Q3 & Did the learning lead to understanding the vocabulary efficiently? \\
    Q4 & Did the learning motivate you to understand the vocabulary? \\
    Q5 & Please feel free to write your thoughts and impressions. \\
  \hline
  \end{tabular}
\end{table}

\begin{table}[h]
  \centering
  \small
  \caption{List of questions after learning with the baseline system}
  \label{tab:survey3}
  \begin{tabular}{c|p{\dimexpr\columnwidth-4em-2\tabcolsep-2\arrayrulewidth\relax}}
  \hline
         & Survey Questions \\ \hline
    Q1 & How difficult was the Pretest? \\
    Q2 & How difficult was the Posttest? \\
    Q3 & Did the learning lead to understanding the vocabulary efficiently? \\
    Q4 & Did the learning motivate you to understand the vocabulary? \\
    Q5 & Please feel free to write your thoughts and impressions. \\
  \hline
  \end{tabular}
\end{table}

\begin{table}[h]
  \centering
  \small
  \caption{List of questions after solving Posttest-2, Posttest-3}
  \label{tab:survey4}
  \begin{tabular}{c|p{\dimexpr\columnwidth-4em-2\tabcolsep-2\arrayrulewidth\relax}}
  \hline
         & Survey Questions \\ \hline
    Q1 & How difficult was the first Posttest? \\
    Q2 & How difficult was the second Posttest? \\
    Q3 & Was first or the second Posttest easier to solve? \\
    Q4 & Please explain why you think so. \\
    Q5 & Please feel free to write your thoughts and impressions. \\
    Q6 & How often do you use ChatGPT? (only asked after Posttest-3) \\
  \hline
  \end{tabular}
\end{table}

\newpage
\section{Sample output of duplicated questions from LLMs}
\label{appendix:badExample}

\begin{table}[h]
    \small
    \centering
    \caption{Example of LLMs not working well as a question-generation system. Similar questions are being asked repeatedly.}
    \label{tab:badExample}
    \renewcommand{\arraystretch}{1.5}
    \begin{tabular}{@{}p{12em}|p{\dimexpr\columnwidth-12em-2\tabcolsep-2\arrayrulewidth\relax}@{}}
    \hline
        Vocabulary to study & overthrow \\ \hline
        Wrong Sentence & The chef decided to overthrow the ingredients into the pot, creating a delicious soup that everyone enjoyed at the dinner party. \\ \hline
        Correct Words & throw, add, mix \\ \hline
        Question from LLMs (1) & Please explain the reasons for the corrections (even if you could not make corrections, please explain based on the answers). \\ \hline
        Answer from the learner (1) & The word ``overthrow'' means to turn something over, but the sentence on the left does not mean to turn something over, but rather to simply put ingredients into a pot. \\ \hline
        Question from LLMs (2) & I see. Which word should I use instead of ``overthrow''? \\ \hline
        Answer from the learner (2) & Words such as `throw', `add', and `mix' are included. \\ \hline
        Question from LLMs (3) & Okay, why is ``overthrow'' wrong? \\ \hline
        Answer from the learner (3) & Because the word ``overthrow'' means to turn something over (to change the orientation of something that was previously in a different orientation). \\ \hline
        Question from LLMs (4) & I see, what other verbs can be used instead of ``overthrow''? \\ \hline
    \end{tabular}
\end{table}

\clearpage
\section{User interface of learning by teaching application}

\begin{figure}[h]
    \centering
    \includegraphics[width=\columnwidth]{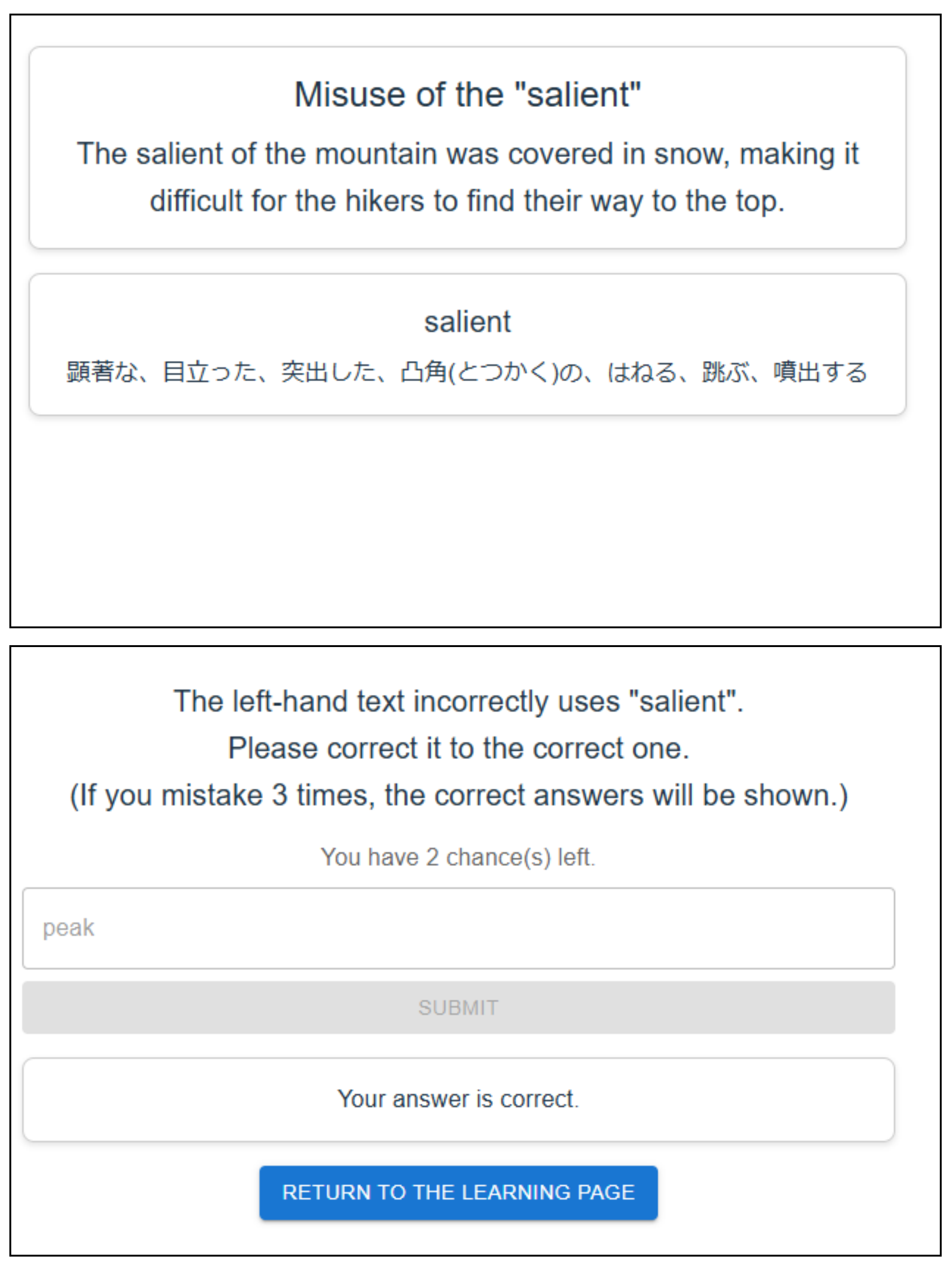}
    \caption{Baseline System: Learning without LLMs.}
    \label{fig:baseline}
\end{figure}

\begin{figure}[h]
    \centering
    \includegraphics[width=\columnwidth]{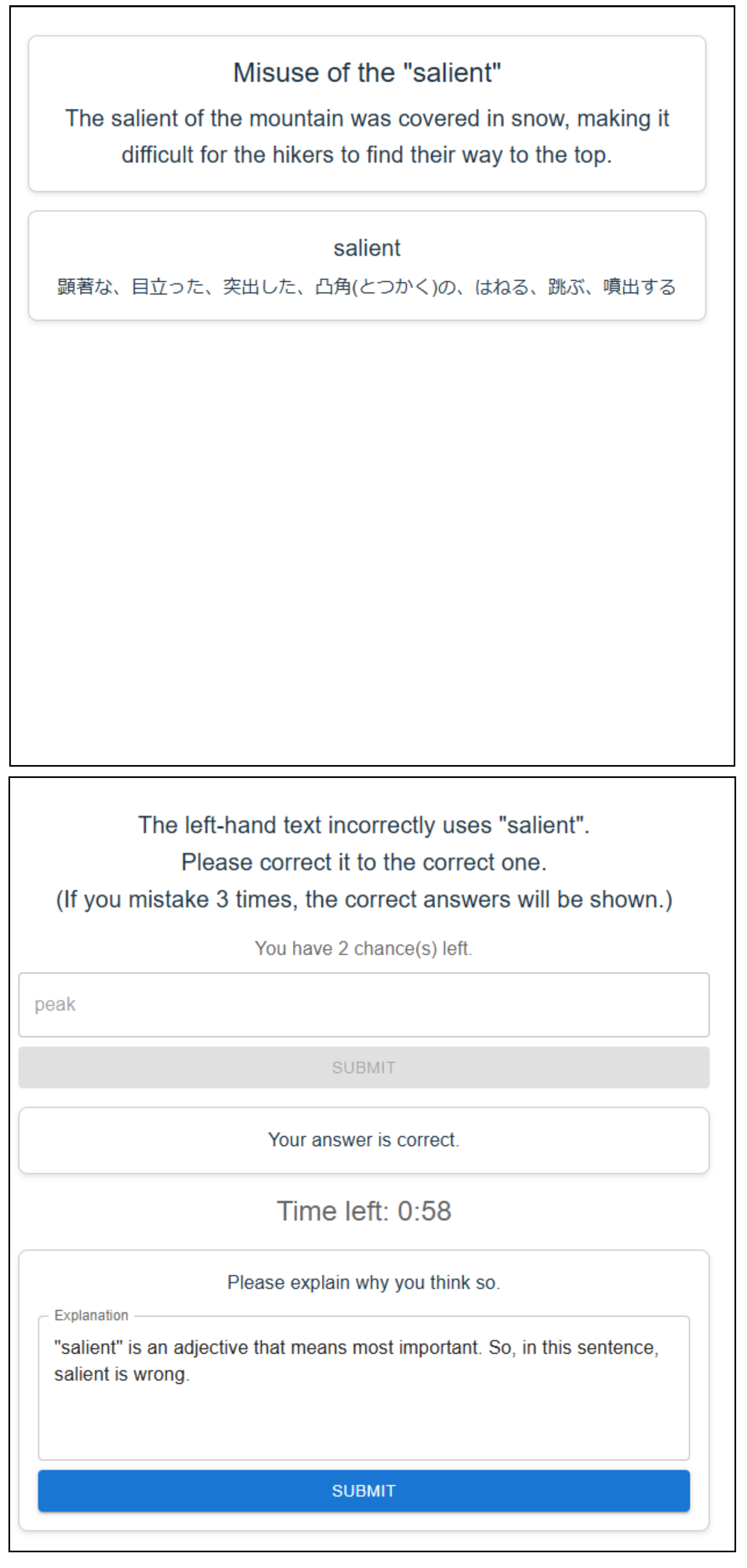}
    \caption{Proposed System: Learning with LLMs.}
    \label{fig:proposed}
\end{figure}

\clearpage
\section{User interface of the pretest questions}
\label{appendix:pretest}

\begin{figure}[h]
  \begin{tabular}{cc}
    \begin{minipage}{0.875\columnwidth}
      \centering
      \includegraphics[width=\columnwidth]{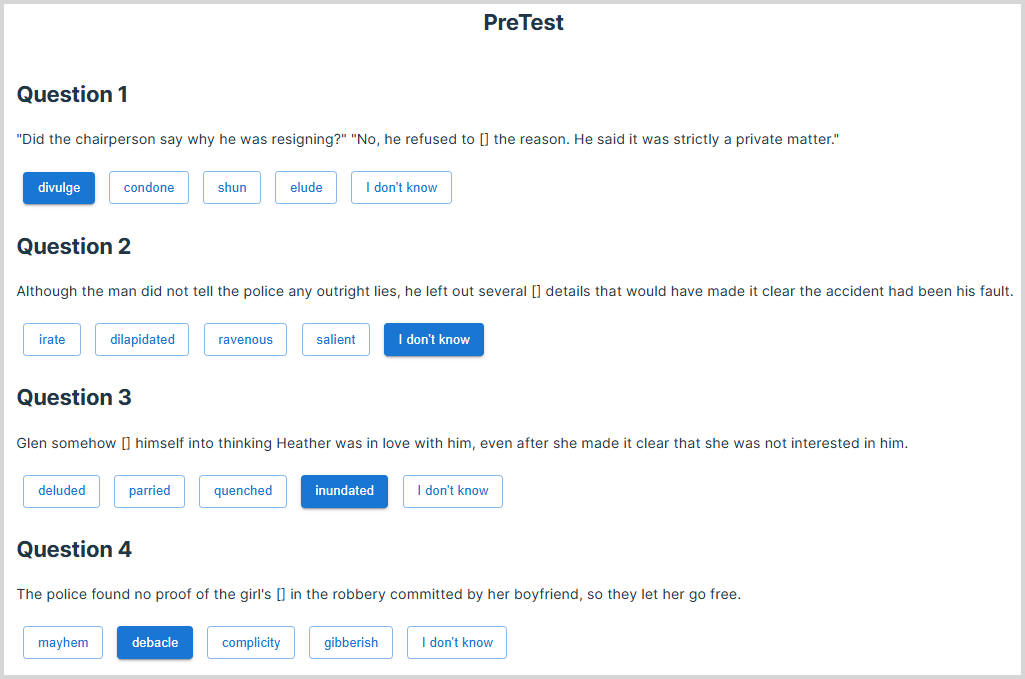}
    \end{minipage}
    \\
    \begin{minipage}{0.9\columnwidth}
      \centering
      \includegraphics[width=\columnwidth]{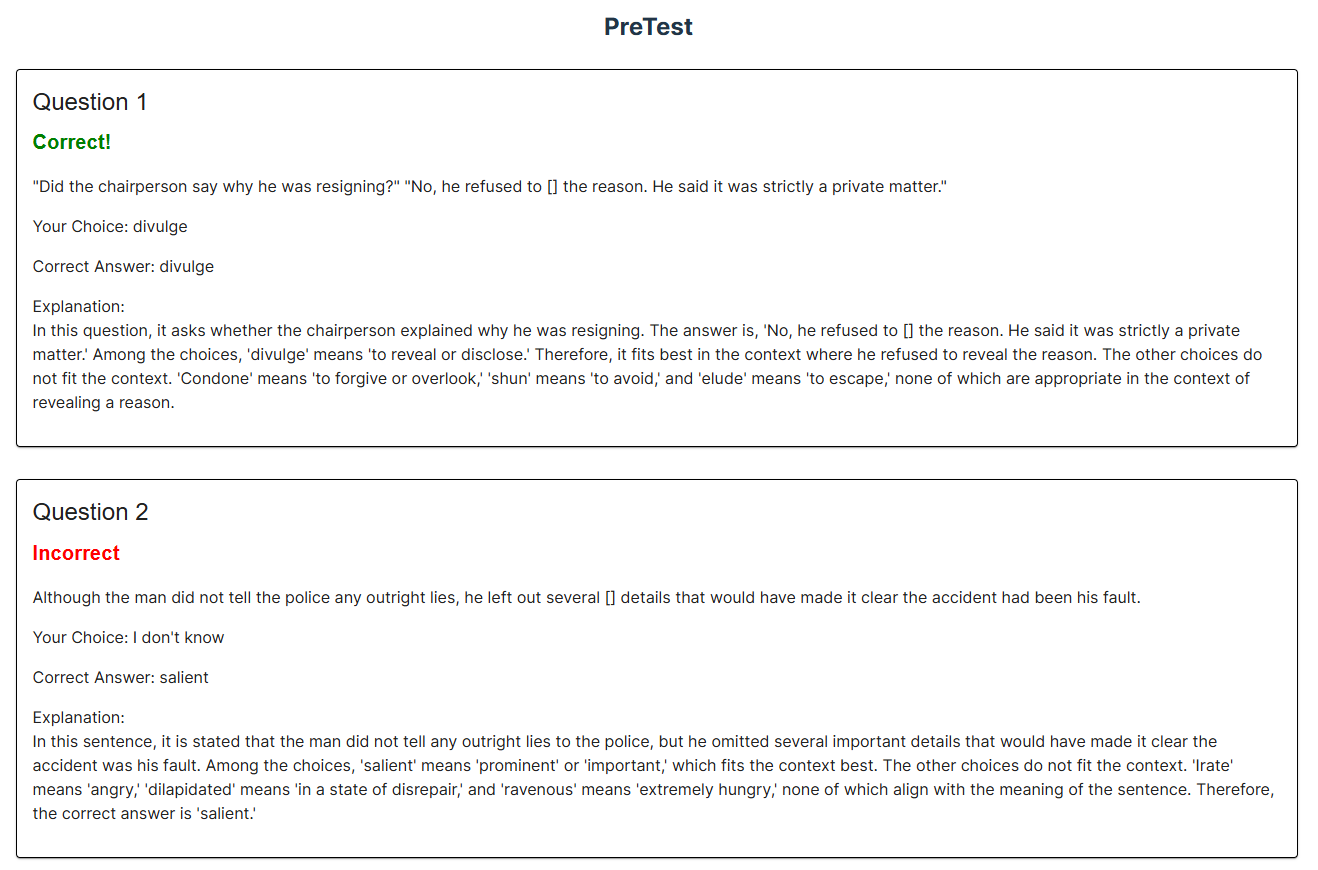}
    \end{minipage}
  \end{tabular}
  \caption{User interface of the pretest questions. Users select multiple-choice questions and receive results after answering questions.}
  \label{fig:test}
\end{figure}

\section{Average number of words entered per interaction}
\label{appendix:inquiryWordsAverage}

\begin{figure}[h]
    \centering
    \includegraphics[width=\linewidth]{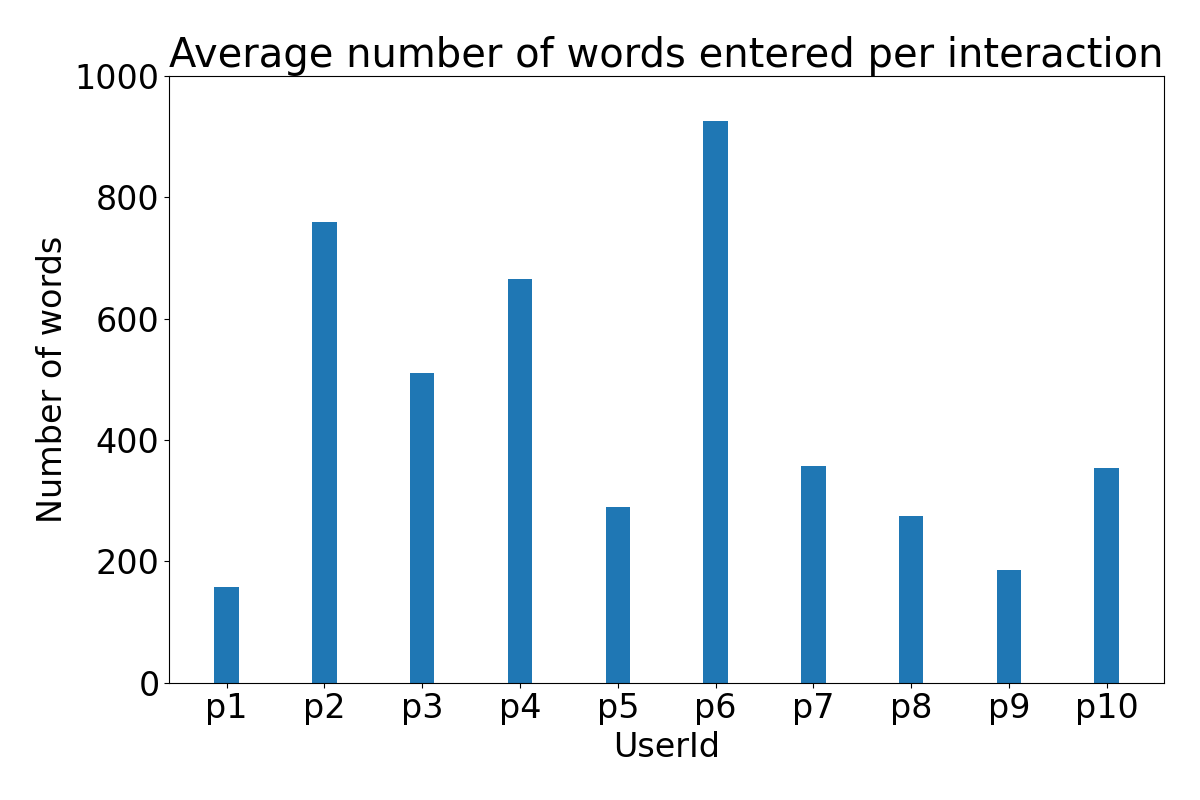}
    \caption{The average number of words entered per interaction by participants. The fewer words, the less interaction between a GPT.}
    \label{fig:inquiryWordsAverage}
\end{figure}

\end{document}